# FLUID CLOCKS:
## Emergence of quantum theory from sub-quantum dynamics

Gerhard Grössing,

*Austrian Institute for Nonlinear Studies,*

Parkgasse 9, A-1030 Vienna, Austria
e-mail: ains@chello.at

**Abstract:** It is argued that any operational measure of time is inseparably bound to the presence of a periodic process in some medium. Since, as first formulated by Einstein's (1905) equation for the energy, all "particles" (neutrons, electrons, photons, etc.) are each characterized by a specific "frequency ", the inverse of this frequency is the smallest operational unit of time available *in principle*. With a corresponding "coarse graining" of an otherwise practically idealized, continuous time (i.e., with the latter then holding on time scales much larger than the coarse grained one), one can show that the basic features of quantum theory can be derived from a minimal set of assumptions. In particular, it is shown here how the Schrödinger equation can be *derived* from classical physics modified only by said assumption of a coarse grained time and the presence of so-called "zero-point fluctuations". The latter relate to the dynamics of a "fluid" vacuum, which is today known to be a far cry from just being "empty space". Consequently, it is shown that time and matter, when discussed from a physical point of view, must *by necessity* be considered in one common framework.

## I. Introduction: Fluid Clocks

In April of this year, the first observation ever was reported of the microscopic structure of a crystalline material fluctuating in time. In a diffraction experiment with x-rays, a metal alloy composed of iron and aluminium was investigated.  The diffraction pattern shows fluctuations in time when the x-ray beam is focused to a very small size of a few micrometers. The number of these temporal fluctuations is so small that they now become visible as x-ray intensity fluctuations (i.e., as opposed to a static pattern when the number is much higher). In other words, this gives "clear-cut evidence that temporal structural fluctuations on an atomic scale are present in the crystal".  [Mocuta *et al.* 2005]

Usually, metals or crystals are often evoking the ideal of some kind of "eternal stability", at least in a metaphorical sense. Hardly does one ever consider these solids as objects, which when looked at close enough, reveal a wild sea of fluctuating constituents. Or, considered the other way 'round, it is seldom recognized that stability in our "macroscopic" world is an *emergent phenomenon* based on a very large number of restless constituents. As Nobel laureate Robert Laughlin argues so convincingly in his new book [Laughlin 2005], emergence is a collective effect of a huge number of particles that usually cannot be deduced from the microscopic equations of motion in a rigorous way and that disappears completely when the system is taken apart. In fact, in said book Laughlin also promotes the idea that one can visualize the vacuum (of only apparently "empty space") as analogous to the ground state of a condensed matter system, with ordinary matter being analogous to excited states of this system.

This is particularly interesting for those who are interested in a deeper understanding of why quantum theory looks (and works) the way it does. For this theory, the most fundamental one concerned with experimentally falsifyable statements on the nature of matter's basic constituents, is also characterized by features of "stability", viz., the linearity of its basic equation, the Schrödinger equation, and the validity of the so-called superposition principle. So, the idea is today increasingly being investigated, whether quantum systems might not be emergent phenomena themselves from an hypothesized underlying non-linear sub-quantum dynamics. (For an early paper on this topic, see [Grössing 1989].) In particular, there exists a long tradition of attempts to explain quantum theory on the basis of an assumed fluid substrate, sometimes (again) called an "aether", whose particle-like constituents undergo diffusion-like processes very similar to the familiar Brownian motion of molecules suspended in water. However, so far approaches focusing solely on the particle constituents of the hypothesized sub-quantum medium in a reductionist manner have largely remained unsatisfactory. The idea of emergence playing a key role in various domains of physics thus leads to the suggestion that also the wave-like phenomena of quantum theory might be nothing but emergent ones due to the presence of a huge number of particle-like constituents of the sub-quantum medium.

In other words, only a strictly reductionist view of nature would claim that, "in the end", or at the "deepest level" of explanations possible, matter would behave according to rules applying to their "ultimate building blocks", like, e.g., some kind of "elementary particles".  Actually, a predominant view of quantum theory for many years has been one that favored such an approach, i.e., that matter was ultimately composed of "particles" following specific rules of propagation (viz., Feynman's path integral formalism, for example). However, along with such an approach also came the view that the specific rules just mentioned are, albeit working perfectly well, not understood at all.

So, it is only legitimate to search for different approaches, including ones which do not assume a strictly reductionist view from the start. In fact, what shall be proposed here is a view which maintains that, at least as far as quantum theory is concerned, no single "ultimate" level of explanations in terms of particles is possible, but rather, that nature might be composed such that a complete description of quantum phenomena is only possible when the physics of particles and of waves are combined within one modelling frame. With the (since many years) well-established knowledge that the so-called "vacuum" is not empty at all, but constitutes some kind of "medium" in which "particles" may propagate, one very natural approach to quantum phenomena may in fact be considered under such premises.

As was shown by the work of Nelson [Nelson 1966 and 1985], motion according to the Schrödinger equation can be reduced to a Brownian-type motion on an assumed "fluid" sub-quantum level, where particles undergo collisions in the framework of diffusion processes according to well-known laws of hydrodynamics. However, one of the problems with Nelson's approach is given by the fact that it cannot account for the by now well-established nonlocal features of the theory. (See, however, [Fritsche and Haugk 2003].)  Still, this may very well be due to the fact that also Nelson's approach exclusively considers one "basic level", i.e., the particle dynamics, only. Considering that classical Brownian motion describes the motion of *particles suspended in some medium*, it would only be logical to consider said hypothesized sub-quantum dynamics in such a way that a description of both particles and waves is necessary, the latter actually originating from the particles, i.e., being created by the particles' "wiggly" motions in the "fluid" medium.

Within a thus circumscribed modelling framework, it is in fact possible to show that the essence of quantum theory (i.e., the Schrödinger equation, Heisenberg's uncertainty relations, the superposition principle and Born's Rule for the probability amplitudes) can be derived from classical physics – albeit only under a new perspective of the latter as well. [Grössing 2004, 2005a, 2005b]

In my approach, both the particle positions and the wave configurations represent "hidden (or rather, with Bell, uncontrollable) variables" to be suitably described by some sub-quantum dynamics. This means, among other things, that Schrödinger's wave function $\psi$ does not need a realistic ontology itself (as Schrödinger himself still assumed), but only represents, just like in the orthodox "Copenhagen" interpretation, the totality of our knowledge about the quantum state. [In this sense, an alternative, i.e., the de Broglie-Bohm interpretation, also got stuck „half way", because in it the (unexplained) wave function still has fundamental status, with the „hidden variables" being determined by functions of it.]

It is only the approaches by Nelson, and recently also my own one, which try to provide an *explanation* of the $\psi$ function and the Schrödinger equation describing its temporal evolution, with particular attention being paid in my work to the nonlocal features of the theory. The latter becomes feasible exactly by attributing wave phenomena an equal status with the particle phenomena. Thus, quantum mechanics (as opposed to classical physics) is pictured as describing the motion of particles subject to *fluctuations* stemming either from a) particle-particle collisions on a sub-quantum level, or b) from wave fronts determined by the wave configurations of the surrounding medium. (A very similar approach is also found in [Baker-Jarvis and Kabos 2003].)

One can show that the latter is co-determined by the totality of the experimental setup, and thus is essentially of nonlocal character. Even in the domain of classical physics, then, particles can be imagined to be the origins of such waves, but in this case – just like in a "zooming" out from the observed object – the fluctuations will be, relatively speaking, too small and thus unobservable. The stochastic particle trajectory then looks "smoothed out", and the waves practically are emitted in a centric-symmetric manner (Fig. 1) so that they don't carry away any momentum – in clear contradistinction to the "zoomed in" view of the quantum case, where exactly this is happening and the corresponding fluctuations (or changes of particle momentum, respectively) are responsible for the "quantum phenomena".

Let me start by recalling the famous „light clock", with which one can nicely illustrate Einstein's ideas on the relativity of simultaneity [Einstein 1905a]. Here we consider only the idealized case of a photon bouncing back and forth between two parallel mirrors. Each time, the photon completes a "roundtrip", the clock "ticks". Thus, this clock has an insurmountable limit of its time measuring capability given by the inverse "frequency" of said roundtrip. The best temporal resolution is given by $\Delta t = 1/\Omega$, and this light clock, being a so-called "harmonical oscillator", can in a completely equivalent manner be represented by a fiber-optic light-guide in the form of a ring with radius $r$. The angular frequency is then given by $\Omega = c/r$, where $c$ is the velocity of light in the light clock. (In other words: Although one may consider a hypothetical, continuous process relating to the propagation of the photon, the operationally relevant quantity corresponds to a discretization of time into minimal time steps $\Delta t$.)

Now we remember Einstein's other work from 1905, i.e., on the "light-electrical effect", where for the first time the energy $E$ of a particle is given in its „quantized" form [Einstein 1905b],

$$E = \hbar\omega, \tag{1.1}$$

where $\hbar = h/2\pi$, with $h$ being Planck's quantum of action and $\omega$ the characteristic angular frequency of the photon. However, since the oscillating photon can be described as a „harmonical oscillator", we can now describe the single photon itself as a "miniaturized light clock": Also in this case are we dealing with an insurmountable limitation of temporal resolution, now given by the angular frequency of the photon itself. In other words, the idealized "continuous" time axis must in an operational sense be discretized, so that one has "coarse grained" intervals of minimal temporal resolution $\delta t$, with their inverse given by a fixed quantity, i.e., the angular frequency: $\frac{1}{\delta t} \to \omega$. This will be of great use in the following. Next to equation (1.1) defining the energy we shall have to consider another empirical finding, i.e., that per degree of freedom one has to add an additional amount of energy, the so-called "zero-point energy"

$$\delta E = E_0 = \frac{\hbar \omega}{2}. \tag{1.2}$$

It represents fluctuations of particle energy based on the "restless" environment of the so-called "vacuum", which is why one also speaks of "vacuum fluctuations". They were first measured by [Mulliken 1924], i.e., before today's quantum theory, or the Schrödinger equation, were formalized.

In other words, as our "miniaturized clocks", with their typical angular frequencies, are embedded in a medium characterized by "vacuum fluctuations", one may consider them as "fluid clocks": periodic oscillations in an environment that is ultimately described by the laws of hydrodynamics.

## II. Emergence of quantum theory from sub-quantum dynamics

Equations (1.1) and (1.2) are, together with de Broglie's equation, in complete analogy to (1.1), for the momentum, $\mathbf{p} = \hbar \mathbf{k}$ (with wave number $\mathbf{k}$), the only empirical, or „fundamental", equations necessary for the derivation of the Schrödinger equation, as shall be shown now. In a first step, we shall derive a useful relation from classical statistical mechanics. Although the "particle" (our "fluid clock"), being embedded by the zero-point fluctuations of the surrounding vacuum, will generally be in a thermo-dynamical state off equilibrium, one can nevertheless assume the limiting case of thermodynamic equilibrium exactly for the states fulfilling some extremal condition. This would then be in accordance with our assumption that the "pure" states of quantum theory correspond to emergent eigenstates of an underlying dynamics which more generally would have to be described by non-equilibrium thermodynamics. Still, for the equilibrium scenario between a "particle" and the surrounding vacuum, and using the familiar expression for entropy, $S_e = k \ln P$ (where $k$ is Boltzmann's constant and $P$ denotes the "probability density"), one can write in differential notation:

$$d\ln P = \frac{dQ}{kT} = -\frac{d(\delta E)}{\delta E},\qquad(2.1)$$

where $Q$ and $T$ describe heat exchange and surrounding temperature, respectively, and $\delta E$ describes an "energy bath", which we can now identify with the thermal bath of the zero-point energy. With the limited temporal resolution $\delta t = 1/\omega$ discussed above, i.e., as given by the inverse angular frequency, one obtains with the insertion of equation (1.2) into (2.1) that

$$d\ln P = -\frac{2}{\hbar}d(\delta E)\delta t.\qquad(2.2)$$

Introducing now, for sufficiently large time spans, the transition from the "coarse grained" time to the "practically continuous" time $t$, we get from (2.2):

$$\frac{\partial}{\partial t}\ln P\,dt = -\frac{2}{\hbar}d(\delta E)dt.\qquad(2.3)$$

Now one can introduce the action function $S$ from classical physics, which provides for the energy that $E = -\frac{\partial}{\partial t}S$. Then, of course, one has for the energy fluctuations $\delta E = -\frac{\partial}{\partial t}\delta S$. They are conserved for short (but still much larger than $\delta t = 1/\omega$) periods of time, so that one can then note that $\delta S =$ constant (which trivially also means that $\frac{\partial}{\partial t}d(\delta S) = 0$). Thus we can rewrite equation (2.3):

$$\frac{\partial}{\partial t}\ln P\,dt = \frac{2}{\hbar}d(\delta S).\qquad(2.4)$$

This equation will be very useful, as will be seen shortly. When identical to zero, it tells us that the relative temporal variation of the probability density $P$ is *extremal* and thus in accordance with a *minimal action principle* for the fluctuation $\delta S$.

Now we turn our attention to the "fluid" part of the model. One usually starts off with the so-called "continuity equation", which describes the classical law of the conservation of a "probability current" (i.e., due to the stochastic nature of the participating fluid elements) in so-called "configuration space". With (normalized) probability density $P$, the continuity equation describes the temporal change of said density in a given spatial volume as determined by the flow of the particle current (described by the gradient, i.e., the Nabla operator, $\nabla$) with velocity $\mathbf{v}$:

$$\frac{\partial P}{\partial t} = -\nabla(\mathbf{v}P). \qquad (2.5)$$

Multiplication by $dt$ and division by $P$ then gives

$$\frac{\partial}{\partial t}\ln P\, dt = -\left(\frac{\nabla P}{P}\cdot\mathbf{v} + \nabla\cdot\mathbf{v}\right)dt. \qquad (2.6)$$

Comparison of equations (2.4) and (2.6) thus finally provides

$$\frac{\partial}{\partial t}\ln P\, dt = \frac{2}{\hbar}d(\delta S) = \frac{2}{\hbar}(\delta\mathbf{p}\,d\mathbf{x} - \delta E\,dt). \qquad (2.7)$$

On the r.h.s. of (2.7), we have introduced two quantities representing the fluctuating contributions of the vacuum to momentum and energy of the particle, respectively. Comparing with equation (2.6), they are explicitly given by

$$\delta\mathbf{p} =: \hbar\mathbf{k_u} = \nabla(\delta S) = -\frac{\hbar}{2}\frac{\nabla P}{P}, \quad \delta E = \frac{\hbar}{2}(\nabla\cdot\mathbf{v}). \qquad (2.8)$$

Equation (2.8) thus introduces a new momentum component according to the assumed additional fluctuations, which can be understood partly as due to the influence of Huygens wave fronts (Fig.1, and [Grössing 2004]) such that for the total momentum one now has

$$\begin{aligned}&\mathbf{p}_{\text{tot}} = \hbar\mathbf{k}_{\text{tot}} = \mathbf{p} + \mathbf{p_u}, \\ &\text{where } \mathbf{p} = \hbar\mathbf{k} = \nabla S, \\ &\text{and } \mathbf{p_u} = \hbar\mathbf{k_u} = -\hbar\nabla P/2P = -\hbar\nabla R/R.\end{aligned} \quad (2.9)$$

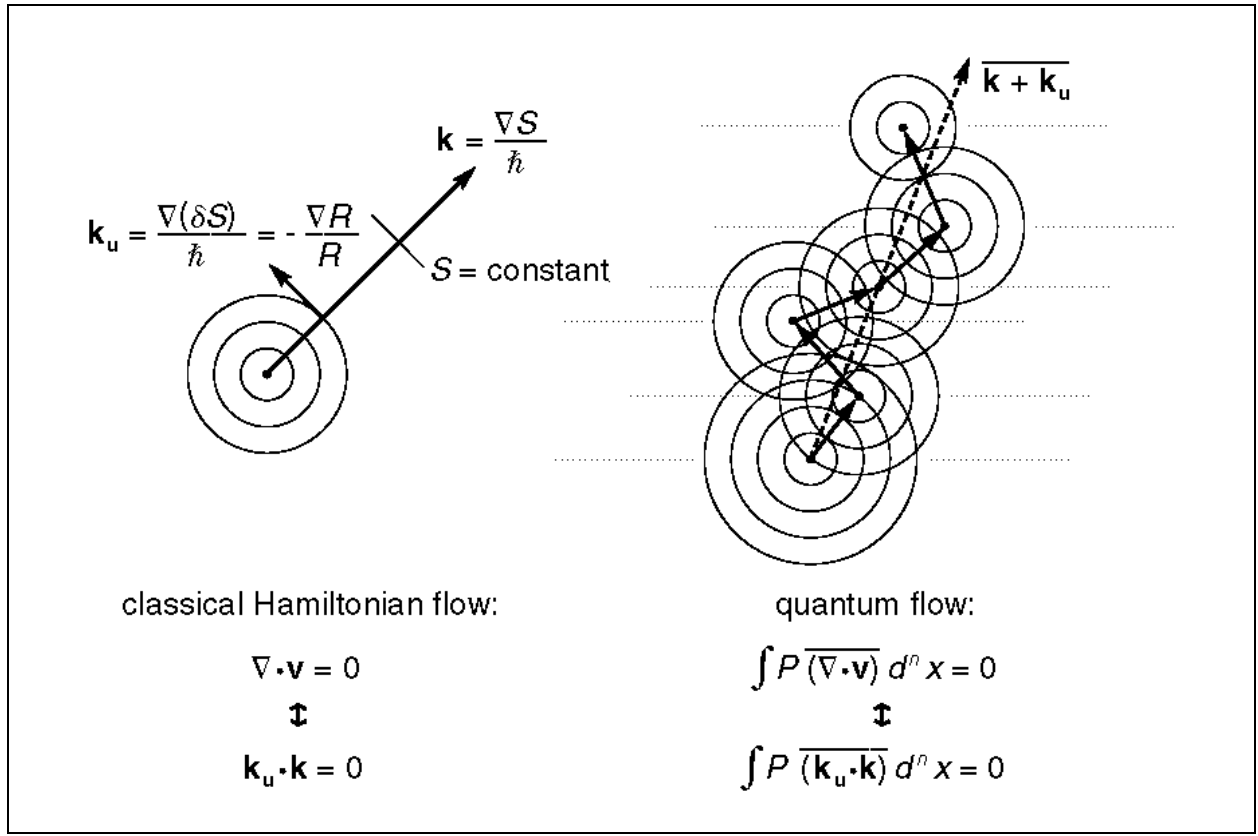

Fig. 1:

Schematic distinction of classical ("Hamiltonian") flow (left) and quantum flow (right), with the circles indicating the propagation of spherical Huygens wave surfaces at arbitrary distances (from [Grössing 2004]). The dotted lines in the figure on the right indicate symbolically that the waves pictured represent only the local surroundings of a generally extending probability field, thus illustrating that the fluctuations shown are to be seen in the context of the whole (and possibly nonlocal) embedding environment.

In equation (2.9) we have also made use of the fact that for the intensity of classical waves with amplitude $R$ it holds that $P = R^2$. When taking the averages (denoted by bars) and integrating over positions and fluctuations, one can show that both the average energy fluctuation and the integral over $\overline{\mathbf{k} \cdot \mathbf{k_u}}$ vanishes (see Fig. 1 for illustration, and [Grössing 2004] for details). One can thus write down the classical action integral with the new momentum expression and with some external potential $V$,

$$A = \int P \left\{ \frac{\partial S}{\partial t} + \frac{\hbar^2 \overline{k_{tot}}^2}{2m} + V \right\} d^n x \, dt, \tag{2.10}$$

where

$$\overline{\hbar \mathbf{k}_{tot}} = \overline{\nabla(S + \delta S)} = \nabla S - \hbar \frac{\overline{\nabla R}}{R} = \hbar \mathbf{k} + \hbar \mathbf{k_u}. \tag{2.11}$$

Now we introduce the so-called "Madelung transformation" (with the star denoting complex conjugation),

$$\psi^{(*)} = R e^{(-)\frac{i}{\hbar} S}. \tag{2.12}$$

Thus one has

$$\frac{\nabla \psi}{\psi} = \frac{\nabla R}{R} + \frac{i}{\hbar} \nabla S, \text{ and } \left| \frac{\nabla \psi}{\psi} \right|^2 = \left( \frac{\nabla R}{R} \right)^2 + \left( \frac{\nabla S}{\hbar} \right)^2, \tag{2.13}$$

and one obtains a *new transformation between the formulations of classical physics and orthodox quantum theory*: the square of the average total momentum is given by

$$\boxed{\overline{p_{tot}}^2 = \hbar^2 \left[ \left( \frac{\nabla R}{R} \right)^2 + \left( \frac{\nabla S}{\hbar} \right)^2 \right] = \hbar^2 \left| \frac{\nabla \psi}{\psi} \right|^2 .} \tag{2.14}$$

With $P = R^2 = |\psi|^2$ from equation (2.12) one can rewrite (2.10) as

$$A = \int L dt = \int d^n x dt \left[ |\psi|^2 \left( \frac{\partial S}{\partial t} + V \right) + \frac{\hbar^2}{2m} |\nabla \psi|^2 \right]. \tag{2.15}$$

Further, with the identity

$$|\psi|^2 \frac{\partial S}{\partial t} = -\frac{i\hbar}{2} \left( \psi^* \dot\psi - \dot\psi^* \psi \right)$$

one finally obtains the well-known "Lagrange density"

$$L = -\frac{i\hbar}{2} \left( \psi^* \dot\psi - \dot\psi^* \psi \right) + \frac{\hbar^2}{2m} \nabla \psi \cdot \nabla \psi^* + V \psi^* \psi. \tag{2.16}$$

As given by the standard procedures of classical physics, this Lagrangian density provides (via the so-called Euler-Lagrange equations) the Schrödinger equation

$$i\hbar \frac{\partial \psi}{\partial t} = \left( -\frac{\hbar^2}{2m} \nabla^2 + V \right) \psi. \tag{2.17}$$

Here, the function $\psi$ does not appear as an operational physical quantity, but only as a *compact notation* which, according to equation (2.14), puts together in one complex-valued quantity the momentum components of waves (fluctuations) and particles. The Schrödinger equation thus provides the temporal evolution of a mathematically compact "wave-particle system", thereby using time $t$ as a parameter only, which is here described as *emergent* from a (yet to be elaborated) sub-quantum dynamics. Among other features, the latter is characterized by time scales $\delta t = 1/\omega$, i.e., by the discretization of the time axis, or, respectively, the presence of "fluid clocks" of frequency $\omega$ in the vacuum-aether.

**Literature:**

Baker-Jarvis, James and Kabos, Pavel 2003, "Modified de Broglie approach applied to the Schrödinger and Klein-Gordon equations", *Physical Review* A 68, 042110

Einstein, Albert 1905a, „Zur Elektrodynamik bewegter Körper", *Annalen der Physik* 17, 891–921

Einstein, Albert 1905b, „Über einen die Erzeugung und Verwandlung des Lichtes betreffenden heuristischen Gesichtspunkt", *Annalen der Physik* 17, 132-148

Fritsche, Lothar and Haugk, Martin 2003, "A new look at the derivation of the Schrödinger equation from Newtonian mechanics", *Annalen der Physik* 12, 6, 371–403

Grössing, Gerhard 1989, "Quantum systems as 'order out of chaos' phenomena", *Il Nuovo Cimento* 103 B, 497-510

Grössing, Gerhard 2004, "From classical Hamiltonian flow to quantum theory: derivation of the Schrödinger equation", in: *Foundations of Physics Letters* 17, 343–362, http://arxiv.org/abs/quant-ph/0311109

Grössing, Gerhard 2005a, "Observing Quantum Systems", *Kybernetes* 34,1/2, 222–240, http://arxiv.org/abs/quant-ph/0404030

Grössing, Gerhard 2005b, "Classical Physics Revisited: Derivation and Explanation of the Quantum Mechanical Superposition Principle and Born's Rule" (preprint) http://arxiv.org/abs/quant-ph/0410236

Laughlin, Robert B. 2005, *A Different Universe*, New York: Basic Books

Mocuta, Cristian, Reichert, Harald, Mecke, Klaus, Dosch, Helmut, and Drakopoulos, Michael, "Scaling in the Time Domain: Universal Dynamics of Order Fluctuations in $Fe_3Al$", *Science*, online Express Reports, 21 April 2005 [DOI: 10.1126/ science. 1110001] http://www.esrf.fr/NewsAndEvents/PressReleases/Living_metals/

Mulliken, Robert S. 1924, "The band spectrum of boron monoxide", *Nature* 114, 349–350

Nelson, Edward 1966, "Derivation of the Schrödinger Equation from Newtonian Mechanics", *Physical Review* 150 (1966) 1079

Nelson, Edward 1985, *Quantum Fluctuations*, Princeton: University Press